\documentclass[aps,prl,10pt,twocolumn,showpacs,superscriptaddress,footnoteinbib]{revtex4}
\usepackage{amsmath}
\usepackage{latexsym}
\usepackage{amssymb}
\usepackage{color}
\usepackage{graphics,epstopdf}
\usepackage{soul}

\begin{document}

\title{ Fine-grained EPR-steering inequalities}

\author{Tanumoy Pramanik}
\email{Pramanik@telecom-paristech.fr}

\author{Marc Kaplan}
\email{kaplan@telecom-paristech.fr} 
\affiliation{LTCI, T\'{e}l\'{e}com ParisTech, 23 avenue dItalie, 75214 Paris CEDEX 13, France}

\author{A. S. Majumdar}
\email{archan@bose.res.in}
\affiliation{S. N. Bose National Centre for Basic Sciences, Salt Lake, Kolkata 700 098, India}

\date{\today}

\begin{abstract}
 
We derive a new steering inequality based on a fine-grained uncertainty relation to capture EPR-steering for bipartite systems.
Our steering inequality improves over previously known ones since it can experimentally detect all steerable two-qubit Werner state with only two measurement settings on each side. According to our inequality, pure entangle states are maximally steerable.
Moreover, by slightly changing the setting, we can express the amount of violation of our inequality as a function of their violation of the CHSH inequality.
Finally, we prove that the amount of violation of our steering inequality is, up to a constant factor, a
lower bound on the key rate of a one-sided device independent quantum key distribution protocol secure against individual attacks.
To show this result, we first derive a monogamy relation for our steering inequality.

\end{abstract}


\maketitle

The development of quantum information led to distinguish three forms of non-local correlations in quantum physics~\cite{EPR, Schrodinger, Bell, Raid_89, Jones07, Buscemi}. 
These are entanglement, steering and Bell non-local correlations. Einstein, Podolsky and Rosen (EPR) introduced entangled quantum states 
in an attempt to show the incompleteness of quantum physics known as the EPR paradox~\cite{EPR}. The same year, Schr\"{o}dinger re-expressed the EPR paradox as the possibility of steering (more generally, known as EPR-steering), i.e., {when Alice and Bob share an entangled state, Alice can affect Bob's state throught her own measurement.
More precisely, a state exhibits EPR-steering if it cannot be modeled
as Bob holding an unknown yet definite state, a 
description known as a local hidden state (LHS) model~\cite{Jones07}.
Bell-type inequalities can be used to rule out local hidden variable (LHV) models. Similarly, steering inequalities are used to rule out the existence of LHS model and thus, demonstrate steerability.
}

According to Wiseman, Jones and Doherty, 
the three forms of non-local correlations are also tightly related to the experimental settings required to test them~\cite{Jones07}.
To test entanglement, both parties need to trust that they perform quantum operations and also trust their measurement devices.
In the case of EPR-steering, only one party assumes that he applies a quantum measurement and that his device is not controlled by a third party.
Finally, Bell non-locality can be be tested without assuming quantum theory and trusting measurement devices.
This leads to a hierarchy in which EPR-steering lies between Bell non-locality and entanglement.


Experimental demonstration of Bell's non-locality has been achieved by several experiments \cite{BI_Exp}. To test EPR-steering, Reid proposed a testable formulation for continuous variable systems based on the position-momentum uncertainty relation~\cite{Raid_89}. 
Denote $(X,P_x)$ and $(Y, P_y)$ the position and corresponding momentum of two correlated modes.
According to the Reid criterion, one needs to infer the uncertainty (measured by the standard deviation) of the quadrature amplitude $X_{\theta k}= \cos[\theta k] X + \sin[\theta k] P_x$ for $k\in\{1,2\}$ from the measurement outcome of the correlated amplitude $Y_{\phi k}= \cos[\phi k] Y + \sin[\phi k] P_y$. 
EPR-steering occurs when
\begin{eqnarray}
(\Delta_{\inf} X_{\theta 1})^2 (\Delta_{\inf} X_{\theta 2})^2 < \frac{1}{4}, 
\label{ReidSt}
\end{eqnarray}
where $\Delta_{\inf} X_{\theta k} = \langle (X_{\theta k} - g_k Y_{\phi k})^2 \rangle $, $k\in\{1,2\}$. The scaling parameter $g_k=\frac{\langle X_{\theta k} Y_{\phi k}\rangle }{\langle Y_{\phi k}^2\rangle}$ arises from imperfect correlation between amplitudes $X_{\theta k}$ and $Y_{\phi k}$ which occurs due to the effect of environmental interaction and finite detector efficiency. The above criterion has been experimentally tested by Ou et al \cite{EPR_St_Exp_C}. 

Since Reid criterion is based on variances, it fails to capture EPR-steering for Bell non-local states whose correlation exists in higher than second 
order~\cite{Chowdhury}. Walborn et al.~\cite{Walborn}  have improved the situation by introducing an entropic steering inequality. 
{
According to this criterion, states admitting LHS models satisfy
\begin{eqnarray}
H({\mathcal P}_B|{\mathcal P}_A) + H({\mathcal Q}_B|{\mathcal Q}_A) \geq \ln \pi e,
\label{EntropicStC}
\end{eqnarray}
where ${\mathcal P}$ and ${\mathcal Q}$ are two non-commuting observables, and subscripts $A$ and $B$ label Alice's observable and Bob's observable, respectively.
The intuition behind Inequality~\ref{EntropicStC} is that if the state has a LHS model, then
Alice's choice of measurement does not affect Bob's state in a way that would violate
the entropic uncertainty relation
\begin{eqnarray}
H({\mathcal P}_B) + H({\mathcal Q}_B) \geq \ln \pi e.
\label{EUR}
\end{eqnarray}
Steering allows to reduce the uncertainty of non-commuting measurements conditioned on Alice's measurement outcome.
The violation of Inequality~\ref{EntropicStC} thus demonstrates EPR-steering.}

For discrete variable systems, EPR-steering theory has been developed by Wiseman, Jones and Doherty~\cite{Jones07}. In their proposal, Alice and Bob both choose observables among $n$ possible ones.
Alice sends $A_k$ to Bob, a random variable that she obtained by operating on her share. Bob then measures the Pauli observable $\hat{\sigma}^B_k$.
For LHS models, the average correlation of outcomes statisfy
\begin{eqnarray}
\frac{1}{n} \displaystyle\sum_{k=1}^n \langle A_k \hat{\sigma}_k \rangle \leq C_n =\max_{A_k} \left( \frac{\lambda_{\max}}{n} \displaystyle\sum_{k=1}^n \langle A_k \hat{\sigma}_k    \rangle \right).
\label{Saund}
\end{eqnarray}
A violation of the above inequality demonstrates that the state shared by Alice and Bob is steerable.
In 2010, based on this criterion, Saunders et al. experimentally demonstrated the steerability of a two qubit Bell-local state (which does not violate any Bell inequality)~\cite{Saunders}.  

{In the present work, we improve the coarse-grained steering criterions of Refs.~\cite{Saunders,Walborn}.
While Reid's criterion was based on Heisenberg's uncertainty relation, Walborn's on entropic uncertainty
relation, our steering inequality is based on fine-grained uncertainty relations (FUR).
Our work is in two parts. We first introduce
a game played between Alice and Bob to characterize steering. We then 
use a FUR to 
upper bound the winning probability when played with states admitting LHS models. This induces
an inequality whose violation is a demonstration of steering.}

In the following, Alice is the supplier of the state and tries to convince Bob that the state she has prepared is steerable.
We consider two different scenarios based on Alice's knowledge about Bob's set of observables before she sends the state. Depending on Alice's knowledge, our steering inequality has two different bounds. Then we discuss the steerability  of pure bipartite entangle states and two qubit Werner states~\cite{Werner_89} given by
\begin{eqnarray}
\rho^W_{AB}= p \rho_S + \frac{1-p}{4} I,
\label{WernerS}
\end{eqnarray}
where $\rho_S$ is the density matrix of  $(|00\rangle_{AB}+|11\rangle_{AB})/\sqrt{2}$.
Using our steering inequality, one can experimentally test the steerability of $\rho^W_{AB}$ for any mixing  parameter chosen from the range $\frac{1}{2} < p \leq 1$, with only two measurement settings for each party. Previously, two measurement settings only allowed demonstration of steerability of $\rho^W_{AB}$  for $p>\frac{1}{\sqrt{2}}$ \cite{Saunders,Caval_09}.

Finally, we study the relation between our inequality and one-sided device independent quantum key distribution (1s-DIQKD)~\cite{tomarenner}.
It is known that getting a positive key rate in a 1s-DIQKD protocol implies the violation of some steering inequality of the state that is used~\cite{Branci}.
We prove that conversely, the violation of our steering inequality implies the security of a certain 1s-DIQKD protocol.
We also prove a quantitative relation between the amount of violation of our inequality and the key rate 
against individual attacks.

Fine-grained uncertainty relations were first introduced by Oppenheim and Wehner \cite{FUR_2}, and later generalized to tripartite systems both in the unbiased \cite{FUR_3} and biased \cite{FUR_Bias} case.
In their work~\cite{FUR_2}, Oppenheim and Wehner show that 
the amount of non-locality measured by the CHSH inequality is bounded by the uncertainty as measured by some FUR.
We extend this approach to steering, showing that the uncertainty between measurement quantified by FURs induces constraints (modeled as a game) on states admitting LHS models. Violation of these constraints thus demonstrates steering.

In the single qubit case, FUR can be described by the following game. Let  Alice receive a binary question $s\in\{0,1\}$ with probability $p(s)=\frac{1}{2}$. 
When she receives the question $s=0$ ($s=1$), Alice measures observable $\sigma_z$ ($\sigma_x$) on the state $\rho_A$. 
She gets outcome $a_s$.
Alice wins the game if she gets spin up outcome (i.e, $a_{s}=0$)
for both question $s=0$ and $s=1$. The winning probability of the above game is given by
\begin{eqnarray}
P_{game} &=& \displaystyle\sum_s p(s) p(a_s=0)_{\rho_A} \nonumber \\
&\leq & P_{game}^{\max} = \max_{\rho_A}P_{game},
\label{WinP}
\end{eqnarray} 
where $p(a_s=0)_{\rho_A}$ is the probability of obtaining spin up outcome for the measurement corresponding to the question $s$ on the state $\rho_A$. $P_{game}^{\max}$ is the maximum winning probability over all possible strategies, i.e., the choice of the single qubit state $\rho_A$ in this game. In the above situation, $P_{game}^{\max}=\frac{1}{2}+\frac{1}{2\sqrt{2}}$ occurs for the eigenstates of $\frac{\sigma_x+\sigma_z}{\sqrt{2}}$, which are known as maximally certain states~\cite{FUR_2}. For the spin down winning condition (i.e, $a_{s}=1$), the maximum winning probability is the same, and achieved using eigenstates of $\frac{\sigma_x-\sigma_z}{\sqrt{2}}$.

Given a game whose winning probability can distinguish the theories satisfying some physical property
from those that don't, the upper bound for theories satisfying this property can sometimes be interpreted as a FUR.
Therefore, measuring the uncertainty arising from incompatible measurements may be sufficient to distinguish 
physical properties.
%
For example, the CHSH winning condition $a\oplus b=s\ t$ \cite{FUR_2} (where $s$ and $t$ are questions given to Alice and Bob, respectively; $a$ and $b$ are their respective answers)
defines a FUR whose violation discriminate non-local theories.
For the winning condition $a\oplus b=1$ (with $s=t$, i.e., both Alice and Bob apply the same measurement)~\cite{FUR_Memo}, $P_{game}^{\max}$ captures the reduction of quantum uncertainty (measured by entropy) in the presence of quantum memory~\cite{Memory}. For tripartite no-signaling correlations,  it is possible to find a game (based on Svetlichny's inequality) that discriminates
classical theories, quantum theories and super quantum correlations~\cite{FUR_3,FUR_Bias}. In all these cases,
the discrimination can be expressed as a FUR.
In our case, we provide a game that distinguishes steering and bound its winning probability for states admitting a LHS model with a FUR. We then bound explicitly the uncertainty measured by the FUR in the case of local hidden states. The result of the combination of both bounds is our steering inequality.


We consider the following game. Alice prepares a large number of copies of a bipartite state $\rho_{AB}$ between systems labeled by `$A$' and `$B$'. She then sends all the systems labeled by `$B$' to Bob. After getting them all, Bob asks Alice to steer each system in the eigenstates of a randomly chosen observable from the set $\{\mathcal{P},\mathcal{Q}\}$. 
Whenever Bob asks to be steered in an eigenstate of $\mathcal P$, Alice applies observable $\mathcal S$ to her system. Similarly, she
applies observable $\mathcal T$ to steer Bob's system to an eigenstate of $\mathcal Q$.
Alice's task is to convice Bob that they share steerable states by communicating her choices of observables and the outcomes.
On other hand, Bob does not trust Alice. He only believes that Alice sent quantum systems and measured them. 
Bob is not convinced by Alice if the correlation of measurement outcomes 
can be described by local hidden state (LHS) model \cite{Jones07}, i.e.,
\begin{eqnarray}
P(a_{\mathcal{A} },b_{\mathcal{B }}) = \displaystyle\sum_\lambda P(\lambda) P(a_{\mathcal{A}}|\lambda) P_Q(b_{\mathcal{B}}|\lambda).
\label{Nst1}
\end{eqnarray}
Here,
$(\mathcal{A}, \mathcal B) \in\{ (\mathcal{S}, \mathcal{P}), (\mathcal T, \mathcal Q)\}$ are the observables.
$a_{\mathcal A}$ and $b_{\mathcal B}$ are Alice's and Bob's measurement outcomes, respectively. 
$P_Q(b_{\mathcal{B}}|\lambda)$ is the probability of obtaining outcome $b_{\mathcal{B}}$ after measuring a quantum system 
specified by the hidden variable $\lambda$.

Using
$\sum_{i} x_iy_i \leq \max_i \{x_i\} \sum_i y_i,$
for $x_i, y_i$ positive, Equation~\ref{Nst1} becomes
\begin{eqnarray}
P(b_{\mathcal{B }}| a_{\mathcal{A}}) \leq \max_{\lambda}[P_Q(b_{\mathcal{B}}|\lambda)]=P_Q(b_{\mathcal{B}}|\lambda_{\max}).
\label{Nst20}
\end{eqnarray}
Since Bob  chooses random observable from~$\{\mathcal{P},\mathcal{Q}\}$, Inequality~\ref{Nst20} becomes
\begin{eqnarray}
\frac{1}{2} P(b_{\mathcal{P}}| a_{\mathcal{S}}) + \frac{1}{2} P(b_{\mathcal{Q}}| a_{\mathcal{T}}) \leq && \max_{\mathcal{P}^*,\mathcal{Q}^*} [\frac{1}{2} P_Q(b_{\mathcal{P}^*}|\lambda_{\max}) \nonumber \\
&& + \frac{1}{2} P_Q(b_{\mathcal{Q}^*}|\lambda_{\max})],
\label{Nst3}
\end{eqnarray}
where $\mathcal{P}^*,\mathcal{Q}^*$ range over all possible maximally incompatible measurements.

{The above inequality is a fine-grained steering criterion satisfied by bipartite states which admit LHS model for the system `$B$'. Its violation for any combination of outcomes $\{a,b\}$ demonstrates steerability. 
In Inequality~\ref{EntropicStC}, the constraints on states admitting LHS models are expressed in terms of average uncertainty where average is taken over all measurement outcomes. In our case, we consider the uncertainty for each particular outcome in a fine-grained way.
Calculating the right-hand side of Inequality~\ref{Nst3} for LHS models gives a steering inequality.
This term measures the uncertainty arising from incompatible measurements $\mathcal P$ and $\mathcal Q$, and
is bounded by the~FUR.}
%

Now we discuss Alice's cheating strategy when $\rho_B$ is a qubit. Alice tries to maximize the left-hand side of Inequality~\ref{Nst3} using a LHS.
We consider two different scenarios.
In \textit{Scenario-I}, Alice gets the description of $\{\mathcal P, \mathcal Q\}$ before sending the states to Bob. Therefore, her whole strategy, including the choice of the state $\rho_B$, depends on the choices of observables.
In \textit{Scenario-II}, Alice prepares the states before getting the description of Bob's observables. She gets this information when the game starts. However, her communication can still depend on Bob's choice of observables.

\textit{Scenario-I :} 
Before sending systems $B$, Alice knows that Bob is going to randomly chose either observable $\sigma_z$ or $\sigma_x$. 
The optimal LHS strategy is the one that maximizes the fine-grained uncertainty relation.
More precisely, depending upon the knowledge of Bob's winning condition and his set of observables, Alice prepares  maximally certain states \citep{FUR_2} which maximize the corresponding winning probability (given by the FUR) and send them to Bob.
For spin up winning condition, Alice prepares all systems in one of the eigenstates of $\frac 1 2 ({\sigma_x+\sigma_z})$ and sends them to Bob.
Then Bob obtains spin up with probability $\frac{1}{2}+\frac{1}{2\sqrt{2}}$~\cite{FUR_2}.
Similarly, Alice prepares eigenstates of $\frac 1 2 ({\sigma_x^B-\sigma_z^B})$ if Bob wins when he gets spin down. This does not change Bob's winning probability.
This is still true if the state of the system $B$ is labeled by a variable $\lambda$ that remains hidden to Bob.  
Using Inequality~\ref{Nst3}, Bob is convinced that $\rho_{AB}$ is steerable only when
\begin{eqnarray}
P(b_{\mathcal{P}}|a_{\mathcal{S}})+P(b_{\mathcal{Q}}|a_{\mathcal{T}}) > 1+\frac{1}{\sqrt{2}}.
\label{Nst41}
\end{eqnarray} 

\textit{Scenario-II :} Here, 
Alice prepares all systems without any knowledge of Bob's set of observables. 
%
In this case, 
we calculate the average winning probability of getting spin up  where average is taken over set of possible obsevables and then maximize it with respect to all possible local hidden states.  
Bob can check that the state violates this maximum and thus conclude that it is steerable.

To calculate the maximum, assume that the hidden state is prepared along  $\hat{n}$ of polar coordinates $\{\theta,\phi\}$ (i.e., $\rho_B=\frac{1}{2}(I+\hat{n}.\vec{\sigma}^B)$), and that the choices of observables are $\mathcal{P}=\hat{p}.\vec{\sigma}$ 
and $\mathcal{Q}=\hat{q}.\vec{\sigma}$, 
where $\hat{p}$ (resp. $\hat{q}$) is the unit vector of polar coordinates $\{\theta_{\hat{p}},\phi_{\hat{p}}\}$ (resp. $\{\theta_{\hat{q}},\phi_{\hat{q}}\}$)
 
 To calculate the average winning probabaility over all possible set of observables for Bob, without loss of generality, we fix observable $\mathcal{P}$ and take average over the observable $\mathcal{Q}$. The average value of the above winning probability is therefore
\begin{eqnarray}
\mathbb E[\frac{1}{2} P_Q(0_{\mathcal{P}})&+& \frac{1}{2} P_Q(0_{\mathcal{Q}})]= \frac{1}{8\pi}\int_{0}^{2 \pi} \int_{0}^{\pi}  \Big(P_Q(0_{\mathcal{P}}) \nonumber \\
&+&P_Q(0_{\mathcal{Q}})\Big) \sin[\theta_{\hat{q}}]  \mathrm{d}\theta_{\hat{q}} \mathrm{d}\phi_{\hat{q}} \nonumber \\
&=& \frac{1}{4} \big(2+\sin(\theta)\sin(\theta_{\hat{q}}) \cos(\phi -\phi_{\hat{q}})\nonumber \\
&&+\cos(\theta) \cos (\theta_{\hat{q}})\big)
\end{eqnarray}
The maximum of this quantity is $\frac{3}{4}$. This is also true if spin down is chosen as winning condition. In this scenario, Inequality~\ref{Nst3} becomes 
\begin{eqnarray}
P(b_{\mathcal{P}}|a_{\mathcal{S}})+P(b_{\mathcal{Q}}|a_{\mathcal{T}}) \leq \frac{3}{2}.
\label{Nst42}
\end{eqnarray}
When Inequality~\ref{Nst42} is violated, the state $\rho_{AB}$ is steerable. 

\textit{Pure entangled state :} Consider that Alice prepares two qubits in the state
\begin{eqnarray}
|\psi\rangle_{AB}=\sqrt{\alpha}~ |00\rangle_{AB} + \sqrt{1-\alpha}~ |11\rangle_{AB}.
\label{Pure}
\end{eqnarray}
When Bob decides to measure $\sigma_z^B$,
Alice makes a spin measurement along the direction $\{\theta_{\hat{s}},\phi_{\hat{s}}\}$, corresponding to the observable $\mathcal{S}$.
Similarly, if Bob measures $\sigma_x^B$, Alice measures along  $\{\theta_{\hat{t}},\phi_{\hat{t}}\}$, corresponding to the observable~$\mathcal{T}$.
When $a=b=0$, the left-hand side of inequality~\ref{Nst3} becomes
\begin{eqnarray}
P(0_{\sigma_z^B}|0_{\mathcal{S}_A})&+&P(0_{\sigma_x^B}|0_{\mathcal{T}_A})= \frac{(4\alpha -1) \cos(\theta_{\hat{s}})+2\alpha +1}{(4\alpha -2) \cos(\theta_{\hat{s}})+2} \nonumber \\
&+&\frac{\sqrt{(1-\alpha)\alpha} \sin(\theta_{\hat{t}}) \cos(\phi_{\hat{t}})}{(2
   \alpha -1) \cos(\theta_{\hat{t}})+1}.
\label{LSHPURE}
\end{eqnarray}
When $\alpha\neq 0$ or $1$, the maximum value of $P(0_{\sigma_z^B}|0_{\mathcal{S}_A}) + P(0_{\sigma_x^B}|0_{\mathcal{T}_A})$ 
is 2. This is achieved for the choices $\theta_{\hat{s}}=\phi_{\hat{s}}=\phi_{\hat{t}}=0$ and $\theta_{\hat{t}}=\arccos(1-2\alpha)$. 
According to our steering test, all pure entangled state are thus maximally steerable: the value of $ P(b_{\mathcal{P}}|a_{\mathcal{S}})+P(b_{\mathcal{Q}}|a_{\mathcal{T}}) $ is equal to its algebraic maximum.

When Alice and Bob both measure either $\sigma_z$ or $\sigma_x$, the left-hand side of Inequality~\ref{Nst3} becomes
\begin{eqnarray}
P(0_{\sigma_z^B}|0_{\sigma_z^A})+P(0_{\sigma_x^B}|0_{\sigma_x^B})= \frac{3}{2}+\sqrt{\alpha (1-\alpha)}.
\end{eqnarray}
According to {\it Scenario-II}, the state $ |\psi\rangle_{AB} $ is steerable for any $\alpha\neq 0$ or 1. 
In this case, the violation of our steering inequality is a function of $\sqrt{\alpha (1-\alpha)}$. Similarly, the violation of CHSH inequality for the state 
$ |\psi\rangle_{AB} $ is given by $2\sqrt{1+4 \alpha (1-\alpha)}$~\cite{Bell_Pure}.
Therfore, this specific choice of measurement allows us connect the CHSH violation with the violation of our steering inequality.


\textit{Werner state :} Here, we consider that Alice and Bob share $\rho^W_{AB}$ (given by Equation~\ref{WernerS}). 
To steer Bob's system in a specific basis, Alice measures the observable corresponding to this basis on her particle, i.e., $\mathcal{P}=\mathcal{S}$ and $\mathcal{Q}=\mathcal{T}$. When Bob chooses his observable from the set $\{\sigma_z^B,\sigma_x^B\}$, for $a=b=0$, the left-hand side of Inequality~\ref{Nst3} becomes 
\begin{eqnarray}
P(0_{\sigma_z^B}|0_{\sigma_z^A})+P(0_{\sigma_x^B}|0_{\sigma_x^A})=1+p,
\end{eqnarray}
where $P(0_{\sigma_z^B},0_{\sigma_z^A})=P(0_{\sigma_x^B},0_{\sigma_x^A})=\frac{1+p}{4}$ and 
$P(0_{\sigma_z^A})=P(0_{\sigma_x^A})=\frac{1}{2}$, and the observables $\sigma_z^A$ and $\sigma_x^A$ are applied to the state $\rho^W_A=Tr_B[\rho^W_{AB}]$. 
The maximum Bell violation of a Werner state is $2\sqrt{2} p$. In {\it Scenario-I}, Werner states are shown to be steerable for $p>\frac{1}{\sqrt{2}}$.
This matches state-of-the-art experiments with two measurement settings~\cite{Saunders,Caval_09}.
In~\cite{Jones07}, it shown how to prove that
Werner states are steerable for $p>\frac{1}{2}$ in the limit of an infinite number of measurement settings.
Using our inequality in {\it Scenario-II}, Werner states are shown to be steerable for $p>\frac{1}{2}$.
Formally, the set of measurements is infinite, but only two are chosen by each party. 
Notice that $p> 1/2$ is tight since for $1/3<p \leq 1/2$, Werner states are entangled but not steerable.

We now connect our steering inequality with the secret key rate in 1s-DIQKD, according to {\it Scenario-I}.
First, we show that Equation~\ref{Nst41} satisfies a monogamy relation. Suppose that the state considered is $\rho_{ABC}$ shared by Alice, Bob and Charlie. Bob's measurements settings are still supposed to be $\sigma_z$ and $\sigma_x$.
Suppose that Alice and Bob violate Equation~\ref{Nst41}. Then we show that Bob and Charlie cannot violate the steering inequality.
Denoting $\mathcal T_{A,B}=P(b_{\mathcal{P}}|a_{\mathcal{S}}) + P(b_{\mathcal{Q}}|a_{\mathcal{T}})$ and
$\mathcal T_{B,C} = P(b_{\mathcal{Q}}|c_{\mathcal{T'}})
+ P(b_{\mathcal{P}}|c_{\mathcal{S'}})$, the monogamy relation writes
\begin{equation}
\frac 1 2 (\mathcal T_{A,B} + \mathcal T_{B,C}) \leq 1+\frac{1}{\sqrt{2}}.
\label{monogamy}
\end{equation}

The proof is by contradiction. Assume that 
$\frac 1 2 (\mathcal T_{A,B} + \mathcal T_{B,C}) > 1+\frac{1}{\sqrt{2}}$.
Now, consider the mixed terms 
$P(b_{\mathcal{P}}|a_{\mathcal{S}}) + P(b_{\mathcal{Q}}|c_{\mathcal{T'}})$ and $P(b_{\mathcal{Q}}|a_{\mathcal{T}})
+ P(b_{\mathcal{P}}|c_{\mathcal{S'}})$. Their average value is equal to $\frac 1 2 (\mathcal T_{A,B} + \mathcal T_{B,C})$.
Moreover, one of the terms has to be larger than or equal to their average. Assume without loss of generality the first one is and consider the state obtained by measuring Alice's and Charlie's shares of $\rho_{ABC}$ and obtaining 
$a_{\mathcal{S}}$ and $c_{\mathcal{T'}}$, respectively. This state satisfies
$\frac 1 2 (P(b_{\mathcal{P}})+P(b_{\mathcal{Q}}))>\frac 1 2+\frac{1}{2\sqrt{2}}$, contradicting the bound on the fine grained uncertainty relation discussed earlier.

This relation can be applied to derive a lower bound on the key rate of a 1sDIQKD protocol. We consider an entanglement based protocol, in which Alice and Bob measure a state $\rho$ and post-select on outcome bits for which they chose either measurements $\{\mathcal P, \mathcal S\}$ or $\{\mathcal Q, \mathcal T\}$. Bob's measurements $\mathcal P$ and $\mathcal Q$ are assumed to be maximally non-commuting.
Alice and Bob estimate the violation of the steering inequality, that is the value $k$ such that 
$ \frac 1 2 (P(b_{\mathcal{P}}|a_{\mathcal{S}})+P(b_{\mathcal{Q}}|a_{\mathcal{T}})) = \frac 1 2 +\frac 1 {2\sqrt 2} + k $. Then,
from Equation~\ref{monogamy}, $\frac 1 2 (P(b_{\mathcal{P}}|c_{\mathcal{S'}})+P(b_{\mathcal{Q}}|c_{\mathcal{T'}}))\leq \frac 1 2+\frac{1}{2\sqrt{2}} -k$.

These bounds immediately translate into bounds on the key rate of the protocol. Denote the random variable representing Alice's, Bob's and Charlie's outcome bits by $A$, $B$, and $C$. Then the key rate $r = I(B:A) - I(B:C)$~\cite{CsiszarKorner} then satisfies $r  \geq \log [({\frac 1 2 +\frac 1 {2 \sqrt 2} +k })/ ({\frac 1 2 +\frac 1 {2 \sqrt 2} -k })]$. The first order approximation finally leads to the following lower bound on the key rate
$$r \geq  \frac{8k}{(2+\sqrt 2)\log 2}.$$
Notice that this bound is linear in the amount of violation of the steering inequality.
For maximum violation, the key rate is $0.5$ and the above lower bound is $0.47$.
In comparison, a similar approach by Paw\l owski and Brunner led to a key rate of
0.0581~\cite{PawloBrunner}.

{To summarize, we derive a new steering inequality based on fine-grained uncertainty relations. 
In Ref.~ \cite{Saunders}, the authors consider the maximum of average correlation of joint measurements in a LHS model, over all possible combinations of outcomes. In Ref.~\cite{Walborn}, the authors consider the minimum of Bob's conditional entropy in a LHS model, where the condition is Alice's communicated outcome. Here, we consider the maximum conditional probability distribution in LHS models, where the condition is again Alice's outcome. Our inequality generalizes both previous works.}
In the derived inequality, we considers only the sum of uncertainties of a particular measurement outcome for the measurement of two different observables.
Hence, we don't require the probability distribution of all possible permutation of measurement outcomes, as described in~\cite{Walborn}.
According to our steering inequality, all pure entangled states are maximally steerable.
Moreover, a suitable choice of the setting allows us to connect the violation of our steering inequality 
with its CHSH violation. 
Our steering inequality leads to 
a tight test for Werner states, which may in turn lead to more experimental-friendly settings for demonstrating steering.
We improve over earlier results by reducing the number of measurements from infinity to two.
In particular, with two measurements on each side,
Saunders' steering inequality (Inequality~\ref{Saund}) can only prove that
Werner states are steerable for $p>\frac{1}{\sqrt{2}}$; this result is recovered by our steering inequality in Scenario~I (Inequality~\ref{Nst41}).
With three measurements on each side, Saunders' Inequality can demonstrate steerability for $p> \frac{1}{\sqrt{3}}$,
and with 10 measurement on each side, for 
$p>0.5236$~\cite{Saunders}.
This approach only leads to a tight test for Werner states in the limit of infinitely many measurement settings~\cite{Jones07,Saunders}.
Our steering inequality (Inequality~\ref{Nst42}) detects the steerability of any Werner state with $p>\frac{1}{2}$ with two measurement settings on each side. Any steerable Werner state can thus be detected with our inequality with the minimum possible number of measurement settings.
Finally, based on a monogamy relation of our steering inequality, we have proved that the violation can be used to lower bound the key rate of a 1sDIQKD protocol secure against individual attacks. We leave it as an open problem to extend it to collective attacks.

{\it Acknowledgements:} The authors thank Damian Markham, Eleni Diamanti, Anthony Leverrier and Tom Lawson for suggestions to enrich this work. A.S.M. acknowledges support from the project SR/S2/LOP-08/2013
of DST, India. T.P and M.K. acknowledge financial support from ANR retour des post-doctorants NLQCC (ANR-12-PDOC-0022- 01).

\end{document}